\documentclass[aps, prl, twocolumn, groupedaddress, showpacs, showkeys]{revtex4}
\usepackage{graphicx}
\usepackage{dcolumn}

\def\ga{\gamma}

\def\si{\sigma}

\def\De{\Delta}

\def\Om{\Omega}

\newcommand{\ben}{\begin{equation}}
\newcommand{\een}{\end{equation}}
\newcommand{\bea}{\begin{eqnarray}}
\newcommand{\eea}{\end{eqnarray}}
\newcommand{\ba}{\begin{array}}
\newcommand{\ea}{\end{array}}
\newcommand{\bit}{\begin{itemize}}
\newcommand{\eit}{\end{itemize}}

\def\math{\mathsurround 0pt}
\def\oversim#1#2{\lower.5pt\vbox{\baselineskip0pt \lineskip-.5pt
        \ialign{$\math#1\hfil##\hfil$\crcr#2\crcr{\scriptstyle\sim}\crcr}}}

\def\half{\frac{1}{2}}

\newcommand{\vev}[1]{\left\langle#1\right\rangle}
\newcommand{\vmol}{v_{\rm{M{\o}l}}}

\begin{document}

\title{\vspace{-12pt}WIMP Dark Matter and the QCD Equation of State}
\author{Mark Hindmarsh$^1$}
\email{m.b.hindmarsh@sussex.ac.uk}
\author{Owe Philipsen$^{1,2}$}
\email{ophil@uni-muenster.de}
\affiliation{$^1$Department of Physics \& Astronomy, University of Sussex, Brighton BN1 9QH, UK\\
                   $^2$Institut f\"ur Theoretische Physik, Westf\"alische Wilhelms-Universit\"at M\"unster,
                   48149 M\"unster,Germany }
\date{\today}

\begin{abstract}
Weakly Interacting Massive Particles (WIMPs) of mass $m$ freeze out at a temperature 
$T_f \simeq m/25$, i.e.~in the range 400 MeV -- 40 GeV for a particle in the typical 
mass range 10 -- 1000 GeV.  The WIMP relic density, which depends on the effective number of relativistic degrees of freedom at $T_f$, may be measured to better than 1\% by Planck, warranting comparable theoretical precision. 
Recent theoretical and experimental advances in the understanding of 
high temperature QCD show that the quark gluon plasma departs significantly from ideal behaviour up to temperatures of several GeV, 
necessitating an improvement of the cosmological equation
of state over those currently used. 
We discuss how this increases
the relic density by approximately 1.5 -- 3.5\% in benchmark mSUGRA models, 
with an uncertainly in the QCD corrections of 0.5 -- 1 \%.
We point out what further work is required to achieve a theoretical accuracy
comparable with the expected observational precision, and speculate that 
the effective number of degrees of freedom at $T_f$ may become measurable in the 
foreseeable future.
\end{abstract}

{MS-TP-05-1}
\keywords{cosmology; Dark Matter; QCD; Quark Gluon Plasma}
\pacs{95.30.+d,12.38.Mh}

\maketitle


A leading candidate for the primary constituent of dark matter is the
Weakly Interacting Massive Particle (WIMP, see e.g.\ \cite{Bertone:2004pz}).
WIMPs of mass $m$ generically freeze out at a temperature of 
around $T_f \simeq m/25$ GeV, and so for typical masses in the range 
10 -- 1000 GeV, $T_f$ lies between 400 MeV and 40 GeV.
The equation governing the relic density depends on both 
energy and entropy densities, and so the WIMP relic density is sensitive to 
the equation of state of the Universe in this temperature range.
It is normally supposed that, 
above the QCD confinement critical 
temperature of $T_c \sim 200$ MeV, the plasma is weakly interacting because of 
asymptotic freedom, and can be treated as an ideal gas, and this assumption is 
built into at least two of the best-known SUSY WIMP relic density packages, 
DarkSUSY \cite{Gondolo:2004sc} and MicrOMEGAs \cite{Belanger:2002nx}. 

However, intensive non-perturbative studies and experiments revealed that 
the high-temperature QCD plasma still departs significantly from an ideal gas 
at temperatures several orders of magnitude higher than $T_c$.  It is therefore 
worth re-examining the equation of state for the Universe in this critical regime,  
and qualitative investigations have been made before \cite{Abazajian:2002yz}.
Using recent progress in both the deconfined \cite{Kajantie:2002wa} and confined
\cite{Karsch:2003vd} phases, 
we construct an improved equation of state and 
investigate the consequences for the WIMP relic density in one of the dark matter 
packages, DarkSUSY. 
DarkSUSY currently uses an equation of state \cite{Srednicki:1988ce} 
based on work in Refs.\ \cite{Olive:1981dy,Olive:1981wz}.
Replacing it with the improved version  
gives upward corrections of over 2\% to the relic densities in a set of 
benchmark mSUGRA models \cite{Battaglia:2003ab}, 
and larger effects for models with lower $T_f$. 

Although small, this correction is becoming significant in the new era of precision 
cosmology: the cold dark  
matter (CDM) density is determined to better than 10\% in single-field inflation models
from a combination of the Cosmic Microwave Background (CMB) angular power spectrum 
and observations of the galaxy power spectrum by 2dF 
\cite{Peiris:2003ff} and by SDSS \cite{Tegmark:2003ud}. 
Planck promises to do much better, with one estimate 
giving a determination to better than 1\% \cite{Balbi:2003en}.

The calculations we use still involve simplifications and systematic errors,
forcing us to model $p(T)$ to some extent, as described  
below. Estimating our systematic errors on the 
QCD equation of state to be on the order of 10\% near $T_c$, we show how
they propagate into uncertainties in the range of 0.5 -- 1\% in the relic density of WIMPs.
Thus, the quest for new physics in observational cosmology depends on further 
improvements in the  accuracy of the 
quantitative understanding of the QCD equation of state.


There has been much effort in calculating the  pressure $p(T)$  of an 
SU($N_c$) gauge theory with $N_f$ 
fundamental fermions at temperature $T$, as reviewed in 
Refs.\ \cite{Blaizot:2003tw,Kraemmer:2003gd,Karsch:2001cy,Laermann:2003cv}.
Perturbative expansions in the coupling constant $g$ of quantities such as the 
pressure converge badly, and particularly for a strongly-coupled theory like 
QCD.  Strictly perturbative expansions, even when expanded 
to O($g^6\ln g$) \cite{Kajantie:2002wa}, seem to 
converge well only at remarkably high temperatures, above $10^5 T_c$,  in sharp 
contradiction to the ideal gas assumption usually made in cosmology.

In the high temperature regime,
progress has been made using perturbative finite-temperature dimensional reduction (DR)
(see \cite{Kajantie:2002wa} and references therein).  
By constructing a sequence of effective theories for the scales 
$2\pi T$, $gT$ and $g^2T$, the last of which has to be treated non-perturbatively, one can get results for arbitrary $N_f$ of massless quarks applicable down to a few times $T_c$.
By fitting for the non-perturbative and as yet unknown $O(g^6)$ coefficient,
the authors of Ref.\ \cite{Kajantie:2002wa} 
were able to match their calculated pressure reasonably well to pure-glue lattice data 
near the critical temperature. 

Around the transition, there now exist lattice calculations for the pressure and energy density 
for $N_f=0, 2, 3$ degenerate quark flavours,  as well as first data for $N_f= 2+1$, i.e.~two light and one heavier flavour.
The pseudo-critical temperatures $T_c(N_f)$ (defined as the peak of a susceptibility) 
are  currently given as $T_c(0) = 271\pm 2$ MeV, $T_c(2) = 173\pm 8$ MeV and 
$T_c(3) = 154 \pm 8$ MeV \cite{Karsch:2000ps}.
It has to be stressed, however, that
only the pure glue case has been extrapolated 
to the continuum. Based on experience with that theory, the dynamical fermion results are estimated to display a systematic error of about 15\% at the currently available lattice spacing.   
Corrections due to quark masses deviating from the physical ones appear to be negligible in comparison \cite{Karsch:2000ps}.

Finally, below the phase transition the hadronic resonance gas model, which treats the plasma as an ideal gas of mesons, baryons and 
their excited states,  matches reasonably well to lattice data \cite{Karsch:2003vd}.


We now review relic density calculations. 
Consider a particle of mass $m$ and number density $n$, undergoing 
annihilations $XX \to \ldots$ with total cross-section 
$\sigma$, assumed to be a typical weak interaction cross-section, proportional 
to  $G_F^2$. Then \cite{Gondolo:1991dk}
\ben
\dot n + 3Hn = -\vev{\si \vmol} (n^2 - n^2_{\rm eq}),
\een
where $n_{\rm eq}$ is the equilibrium number density, $H=\dot a/a$ is the Hubble parameter, 
and $\vmol$ is the M{\o}ller velocity
which is a relativistic generalisation of the relative velocity of the annihilating 
particles \cite{Gondolo:1991dk}.

In order to solve the equation it is convenient to convert the time variable to $x = m/T$, and 
to measure the relic abundance in terms of $Y = n/s$, where $s$ is the entropy density. 
If the total entropy $S = sa^3$  is conserved, then we can write 
\ben
\frac{dY}{dx} = \frac{1}{3H}\frac{ds}{dx}\vev{\si \vmol}(Y^2 - Y_{\rm eq}^2).
\een
This adiabaticity assumption is violated if the QCD transition is first order, but it is 
most likely to be a cross-over transition at the low chemical potentials which are 
relevant for the early Universe \cite{Karsch:2001cy,Laermann:2003cv}.

Using the Friedmann equation $H^2 = 8\pi G\rho/3$, and defining effective numbers of 
degrees of freedom for the energy and entropy densities, $s = (\rho + p)/T$, through
\ben
\rho = \frac{\pi^2}{30} g_{\rm eff}(T) T^4, \quad s = \frac{2\pi^2}{45} h_{\rm eff}(T) T^3,
\label{e:geffs}
\een
one finds an approximate solution 
\ben
Y_0 \simeq \left(\frac{45}{\pi}\right)^\half\frac{1}{mM_\mathrm{P}\vev{\si \vmol}_{T_f}}\frac{x_f}{g_*^\half(T_f)},
\een
where $T_f$ is the freeze-out temperature, defined to be the temperature at which 
the relic abundance is a certain factor (taken to be 2.5) above the equilibrium abundance.
We see that the relic density depends on the parameter
\ben
g^{1/2}_*(T) = \frac{h_{\rm eff}}{g_{\rm eff}^{1/2}}
 \left(1 + \frac{T}{3} \frac{d \ln h_{\rm eff}} {d T} \right).
\label{e:gstar}
\een
It is through this parameter that the QCD equation of state influences the relic density. 


In an ideal gas at temperature $T$,  
a particle of mass $m_i= x_iT$ contributes to $g_{\rm eff}$, $h_{\rm eff}$ the amounts 
\bea
g_{i,\rm eff} \equiv \frac{\rho_i}{\rho_{0}} &=& 
\frac{15}{\pi^4} \int_{x_i}^\infty \frac{(u^2 - x^2_i)^\half}{e^u \pm 1}u^2du,\\
h_{i,\rm eff} \equiv \frac{s_i}{s_{0}} &=& 
\frac{45}{12\pi^4} \int_{x_i}^\infty \frac{(u^2 - x_i^2)^\half}{e^u \pm 1}(4u^2-x_i^2)du,
\eea
where $\rho_0$ and $s_0$ are the energy and entropy densities for a free massless boson.

Interactions correct the ideal gas result, and $g_{\rm eff}, h_{\rm eff}$ have to be extracted from 
calculations of the energy
and entropy, Eqs.~(\ref{e:geffs}). Note that in the relevant 
temperature range 40 to 0.4 GeV, 16 out the 18 bosonic degrees of freedom are 
coloured, with the coloured fermionic degrees of freedom dropping from 60/78 to 36/50. 
The dominant corrections are therefore expected to come from 
the coloured sector of the Standard Model, weak corrections are moreover suppressed by the boson masses and negligible.

The relic density codes DarkSUSY and MicrOMEGAs 
use identical equations of state, developed in Refs.\  
 \cite{Olive:1981dy,Olive:1981wz,Srednicki:1988ce}. 
Below $T_c$ the hadronic 
degrees of freedom are modelled by an interacting gas of hadrons and their 
resonances, while above $T_c$ the quarks and gluons are taken to 
interact with a linear potential $V_Q(r) = Kr $, with a phenomenologically motivated value
$K = 0.18 \textrm{GeV}^2$, derived from the slope of Regge trajectories.  
In this model, the pressure  is already very close to ideal at temperatures above 1.6 GeV.
All other Standard Model particles are free.

In this work we also take the ideal gas contributions for the particles of the Standard Model, with masses given by the Particle Data Group central values \cite{Eidelman:2004wy}. 
In the confined phase quarks and gluons are replaced by hadronic models described below.
In the deconfined phase, the contribution to the pressure 
of the coloured degrees of freedom is scaled 
by a function $f(T)$, defined to be the ratio between the 
true QCD pressure $p(T)$ and the Stefan-Boltzmann result $p_{\rm SB}$ for the same 
theory, $f(T) = p(T)/p_{\rm SB}(T)$. 
This correction factor is derived from lattice \cite{Karsch:2000ps} and perturbative 
\cite{Kajantie:2002wa} calculations for $N_f=0$, and uses an approximate universality in the pressure curves for different $N_f$ observed by Karsch et al.\ \cite{Karsch:2000ps}. 
Near the transition, the lattice-derived curves for $f(T)$ have the  
approximate form 
\(
f(T,N_f) = f_{\rm QCD}(T/T_c(N_f)),
\) 
where $T_c(N_f)$ is the critical temperature for the theory with $N_f$ light fermion flavours.
Besides this, there appears to be only negligible $N_f$-dependence within the current numerical accuracy. 
We are therefore motivated to neglect quark mass effects, take the $N_f=0$ lattice data (in the continuum limit) and the $N_f=0$ DR formula of Kajantie et al.\  
\cite{Kajantie:2002wa}, and scale the temperature 
dependence by $T_c(3)/T_c(0)$. The correction factors are matched 
at 1.2 GeV using the undetermined O($g^6$) parameter, which we take to be 0.6755, 
close to the value 0.7 used in Ref.\ \cite{Kajantie:2002wa}.
At higher temperatures one crosses the $c$ and $b$ mass thresholds. 
However, for $N_f >3$ the O($g^6$) 
fitting parameter is unknown, and the 
appropriate critical temperatures are also unknown. 
Hence we believe that scaling the $N_f=0$ result is the best we can do 
at present.  We will discuss later how improvements can be made.

In the confined phase we label our model equations of state (EOS) A,B and C.
EOS  A ignores hadrons completely, as
the lattice shows that $f(T)$ very rapidly approaches 
zero below $T_c$.
EOS B and C model hadrons as a gas of free mesons and baryons. 
It was noted in Ref.\ \cite{Karsch:2003vd} that such a gas, 
including all resonances, gives a pressure which fits remarkably well to the $N_f = 2+1$ lattice results although, as the authors themselves point out, this result should be treated with 
caution as the simulations are not at the continuum limit. 
We include all resonances listed in the Particle Data Group's 
table \texttt{mass\_width\_02.mc} \cite{Eidelman:2004wy}.  

We make a sharp switch to the hadronic gas  
at a temperature $T_{\rm HG}$.  For our EOS B we take
 $T_{\rm HG} = T_c =154$ MeV, and for EOS C we take
 $T_{\rm HG} = 200$ MeV and $T_c = 185.5$ MeV, values chosen to give as smooth 
 a curve for $h_{\rm eff}$ as possible.
The effects of these equations of state on the relic densities turn out to differ by less than 0.3\% in the relevant temperature interval, 
so in the following we concentrate on EOS B.

Before presenting our results 
we note that $\Om_ch^2$ is directly proportional to the entropy density today
$s_0 = (2\pi^2/45)h_{\rm eff} (T_\ga)T^3_\ga$, and that this must be determined as accurately as possible.  The photon temperature $T_\ga = 2.725\pm0.001$ is very accurately measured
\cite{1999ApJ...512..511M}, 
but the contribution from neutrinos requires a separate freeze-out calculation. Recent work
\cite{Mangano:2001iu} gives $h_{\rm eff}(T_\ga) = 3.9172$.  We find that taking 
freeze-out temperatures to be 3.5 MeV for $\nu_\mu$ and $\nu_\tau$ and 2 MeV for 
$\nu_e$, as recommended in Ref.\ \cite{Srednicki:1988ce}, gives 
$h_{\rm eff}(T_\ga) = 3.9138$, which is accurate enough for our purposes. 
DarkSUSY uses $h_{\rm eff}(T_\ga) = 3.9139$.

In Figs.\ 1 we plot for our EOS B the effective numbers of degrees of freedom 
$h_{\rm eff}(T)$ and $g^{1/2}_*(T)$ defined 
 in Eqs.\ (\ref{e:geffs}) and (\ref{e:gstar}), compared with those used in 
 DarkSUSY \cite{Gondolo:2004sc} and MicrOMEGAs \cite{Belanger:2002nx}.
The spike in our $g^{1/2}_*(T)$ is an artefact of the matching of the 
scaled lattice data and the hadronic equation of state (i.e.~the first derivative jumps). 
Since for physical QCD the transition is a smooth crossover, this spike is 
unphysical. However, it has no noticeable influence on the freeze-out of WIMPs at 
higher temperatures.

\begin{figure}
\begin{center}
\includegraphics[width=0.94\hsize]{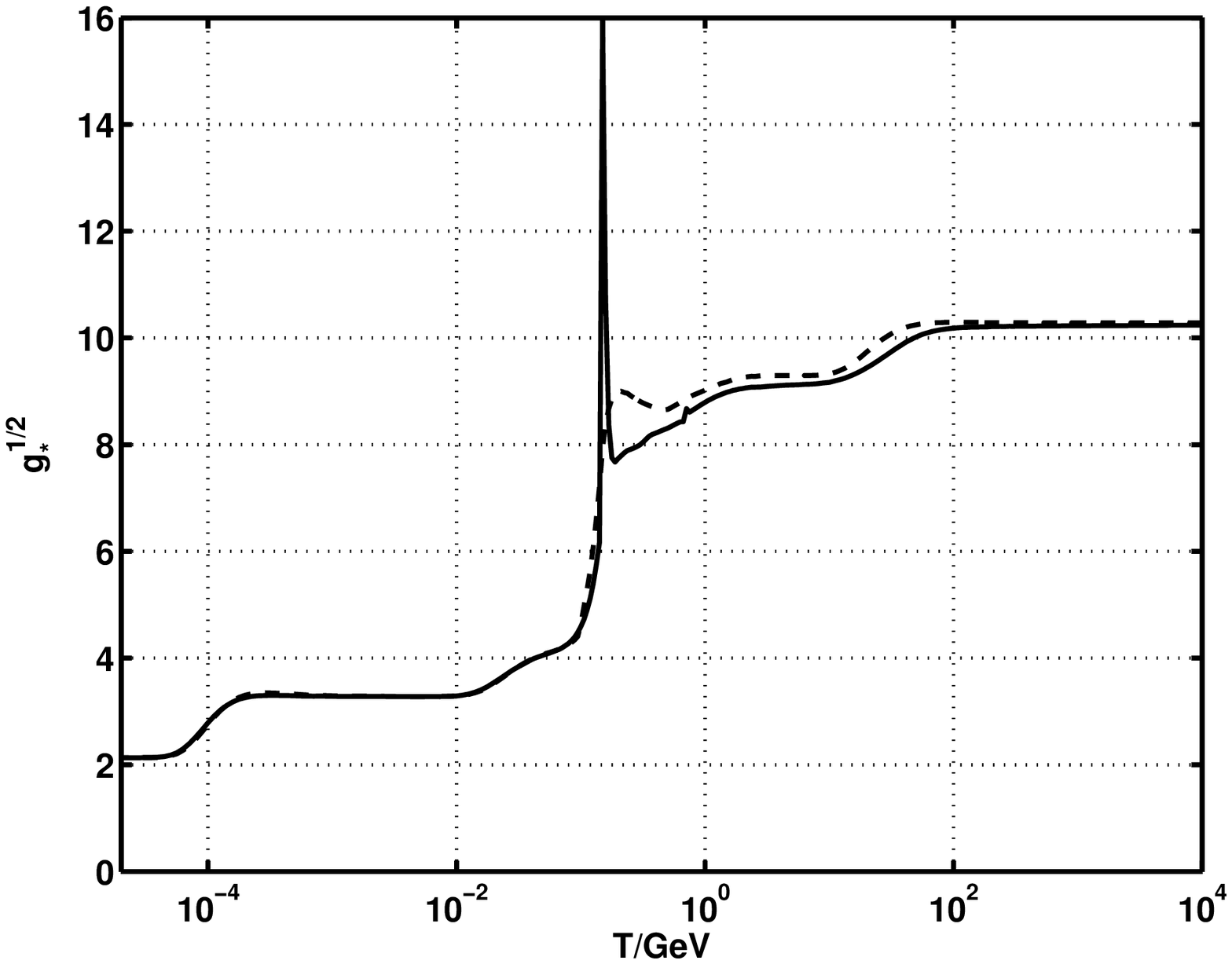}

\includegraphics[width=0.94\hsize]{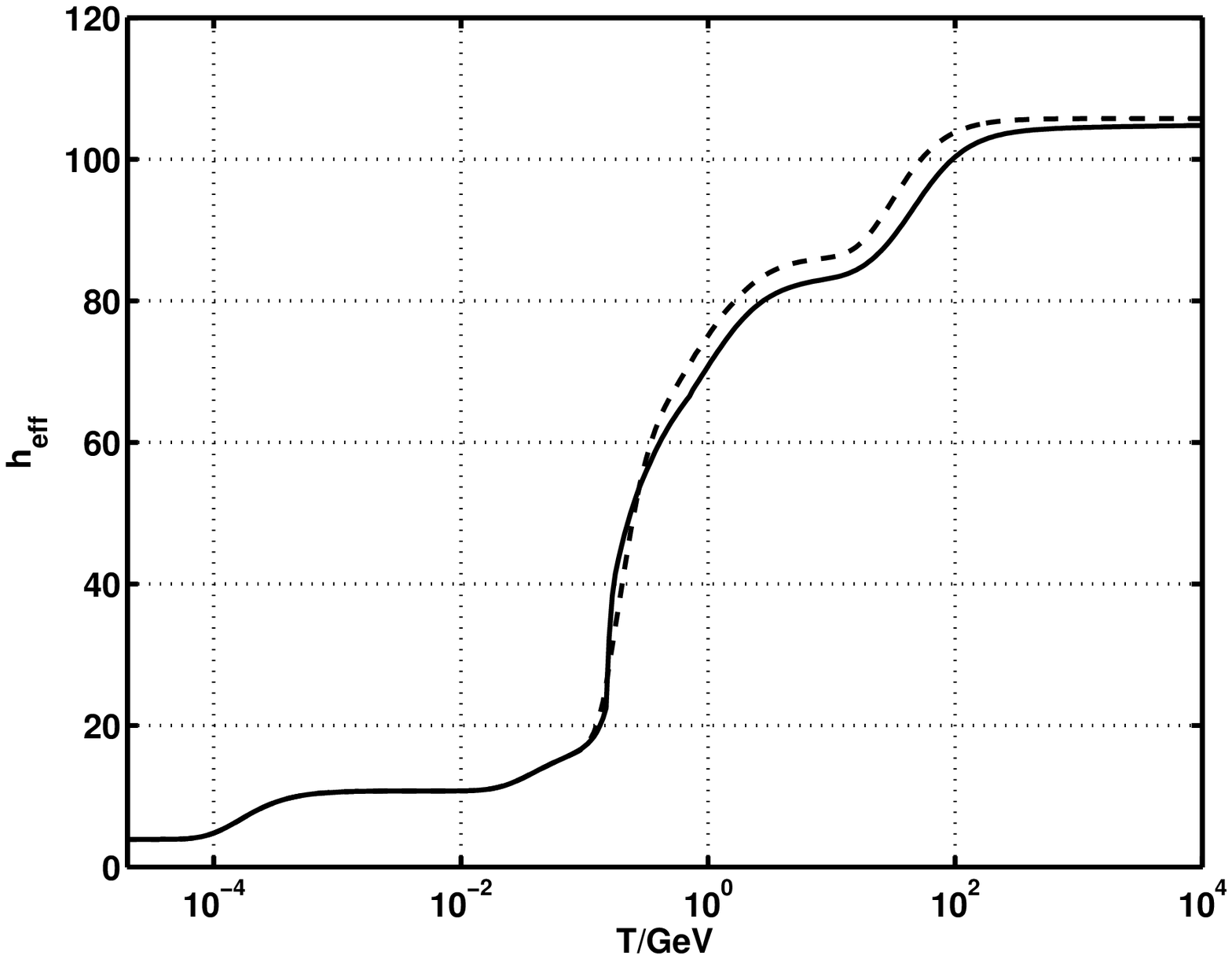}
  \caption{\label{f:dofs} 
  Degrees of freedom factors $g^{1/2}_*(T)$  and $h_{\rm eff}(T)$, defined 
  in Eqs.\ \ref{e:geffs} and \ref{e:gstar}, for our equation of state B (solid) compared with 
  those currently used in DarkSUSY and MicrOMEGAs (dashed).
  The spike in the upper panel is a numerical artefact without physical impact, see text.}
\end{center}
\end{figure}

In Table \ref{t:rdens}  we exhibit the effect of the new equation of state on the density of 
relic neutralinos $\chi$, for the mSUGRA models used to test DarkSUSY in the standard 
distribution \cite{Battaglia:2004mp}.  
We find changes of  about 1.5--3.5\%. 
In order to quantify the effect of uncertainty in the lattice data, we introduce 
two new models B2 and B3, which are constructed by scaling the lattice curve by 0.9 and 1.1 respectively, and then adjusting the O($g^6$) parameter in the DR pressure curve so that 
it meets the scaled lattice curve at $T= 4.43T_c$.  
Thus a 10\% uncertainty 
in the lattice pressure curve translates to an uncertainty in the relic density in the range 
 0.5 -- 1\%. Note that the lowest
freeze-out temperature in the table is  about 4 GeV.
This corresponds to more than $20T_c$, where the QCD corrections to $g_*$ are around 5\%. Evidently, WIMPs with freeze-out 
temperatures closer to $T_c$, such as sterile neutrinos \cite{Abazajian:2002yz}, 
would be affected more strongly.
\begin{table}
\begin{tabular}{l|c|c|c|c|c}
Model & $m_\chi$/GeV &  $T_f$/GeV &$\Om_ch^2$ (DS) &$\Om_ch^2$ (here) & $\Delta$(\%) \\
\hline
A'& 242.83 & 9.8 &0.0929 & 0.0948$^{(54)}_{(42)}$ & 2.0$^{2.6}_{1.4}$\\[2pt]
B'& 94.88 & 4.1 & 0.1213  & 0.1242$^{(56)}_{(31)}$ & 2.4$^{3.6}_{1.5}$\\[2pt]
C'& 158.09 & 6.5 & 0.1149  & 0.1174$^{(83)}_{(65)}$& 2.2$^{2.9}_{1.5}$\\[2pt]
D'& 212.42 & 8.6 & 0.0864  & 0.0882$^{(88)}_{(76)}$& 2.0$^{2.7}_{1.4}$\\[2pt]
G'& 147.98 & 6.2 & 0.1294  & 0.1323$^{(33)}_{(13)}$& 2.2$^{3.0}_{1.4}$\\[2pt]
H'& 388.38 & 16.0 & 0.1629  & 0.1662$^{(71)}_{(53)}$& 2.0$^{2.6}_{1.5}$\\[2pt]
I'& 138.08 & 5.8 & 0.1319  & 0.1351$^{(62)}_{(40)}$& 2.4$^{3.2}_{1.6}$\\[2pt]
J'& 309.17 & 12.6 & 0.0966  & 0.0984$^{(90)}_{(79)}$& 2.0$^{2.5}_{1.4}$\\[2pt]
K'& 554.19 & 22.9 & 0.0863  & 0.0883$^{(88)}_{(78)}$& 2.3$^{3.0}_{1.8}$\\[2pt]
L'& 180.99 & 7.5 & 0.0988  & 0.1011$^{(18)}_{(03)}$& 2.3$^{3.0}_{1.5}$\\[2pt]
\hline
\end{tabular}
\caption{\label{t:rdens} Relic densities in 
benchmark models \cite{Battaglia:2003ab}, calculated with DarkSUSY.  
Displayed: 
neutralino mass $m_\chi$, freeze-out temperature $T_f$, $\Om_ch^2$ computed 
using the current DarkSUSY equation of state, $\Om_ch^2$ using our QCD-corrected Equations of State B$^{\rm(B2)}_{\rm(B3)}$, and percentage changes $\De$.}
\end{table}


To conclude: by updating the equation of state of the Standard Model in the light of recent 
developments in high temperature QCD, we have found differences in WIMP 
relic density calculations of a few per cent for the benchmark models 
of Ref.\ \cite{Battaglia:2003ab},  a small but not insignificant effect.
These models freeze out well above $T_c$, 
and so it is likely that there are other models 
exhibiting greater changes.  
In single field inflation models,  
the 68\% confidence limit on the dark matter density is already $\Om_c h^2 = 0.127\pm 0.017$
from the WMAP First Year data \cite{Peiris:2003ff}, while Planck has been estimated 
to be able to reach a level of $\De\Om_c h^2 = 0.0011$ \cite{Balbi:2003en}. 
In order to benefit from this accuracy, it has to be matched
by that of theoretical calculations of the QCD equation of state.  
To obtain an accuracy of 1\% we estimate that we require lattice data accurate to 10\% 
at about $T \simeq 4T_c$, for realistic quark masses, and in the continuum limit.  
Further progress is possible: the O($g^6$) term is in principle calculable 
but requires a 4-loop computation in lattice perturbation theory \cite{Kajantie:2002wa}.
Moreover, the $b$ and $c$ quark mass thresholds may affect the QCD corrections. 
We also note that the equation of state at a few $T_c$ is also
experimentally accessible in current and future heavy ion experiments, which gives them a 
direct cosmological motivation.  

Finally, given a sufficiently accurate $g*$, we speculate that with a combination of measurements of 
the neutralino mass and cross-sections in direct detection experiments 
and at LHC and the ILC, the effective number of degrees of freedom
in the early Universe at a few GeV will 
become experimentally accessible, motivating attempts to improve further on this work.


\begin{acknowledgments}
We are indebted to F.~Karsch and J.~Engels  for lattice data, to M.~Laine for 
providing details from Ref.\ \cite{Kajantie:2002wa}, 
and to P.~Gondolo and F.~Boudjema for discussions.
We acknowledge support from PPARC (OP, Advanced Fellowship) and from the European Network for Theoretical Astroparticle Physics (ENTApP), member of ILIAS, EC contract number RII-CT-2004-506222 (MH).
\end{acknowledgments}


\bibliography{qcddm}

\end{document}